\documentclass[12pt,draftcls,onecolumn]{IEEEtran}
\ifCLASSINFOpdf
\else
\fi
\usepackage{epsfig}
\usepackage{amssymb}
\usepackage{amsmath}
\usepackage{amsfonts}

\usepackage{graphics}
\usepackage{graphicx}
\usepackage{epsfig}
\usepackage{color}
\usepackage{subfigure}
\usepackage{extarrows} 
\usepackage{stmaryrd}
\usepackage{txfonts}
\usepackage{mathrsfs}
\usepackage{latexsym, bm}
\newtheorem{theorem}{Theorem}

\newtheorem{definition}{Definition}
\newtheorem{lemma}{Lemma}

\newtheorem{remark}{Remark}
\newtheorem{example}{Example}

\newcommand*{\QEDA}{\hfill\ensuremath{\blacksquare}}
 
\DeclareMathOperator{\diag}{diag}

\begin{document}
%

\title{Consensus of Hybrid Multi-agent Systems}
%
%
%

\author{Yuanshi~Zheng,~
        Jingying~Ma,
       and~Long~Wang~
\thanks{This work was supported by NSFC (Grant Nos. 61533001, 61375120 and 61304160) and the Fundamental Research Funds for the Central Universities (Grant No. JB140406).}
\thanks{Y. Zheng and J. Ma are with Key Laboratory of Electronic Equipment Structure Design of Ministry of Education, School of Mechano-electronic Engineering, Xidian University, Xi'an 710071, China and also with the Center for Complex Systems, School of Mechano-electronic Engineering, Xidian University, Xi'an 710071, China
 (e-mail:zhengyuanshi2005@163.com;majy1980@126.com)
 }
\thanks{L. Wang is with the Center for Systems and Control, College of Engineering, Peking
University, Beijing 100871, China
 (e-mail: longwang@pku.edu.cn)}
}

\maketitle

\begin{abstract}
In this paper, we consider the consensus problem of hybrid multi-agent system. First, the hybrid multi-agent system is proposed which is composed of continuous-time and discrete-time dynamic agents. Then, three kinds of consensus protocols are presented for hybrid multi-agent system. The analysis tool developed in this paper is based on the matrix theory and graph theory. With different restrictions of the sampling period, some necessary and sufficient conditions are established for solving the consensus of hybrid multi-agent system. The consensus states are also obtained under different protocols. Finally, simulation examples are provided to demonstrate the effectiveness of our theoretical results.
\end{abstract}


%
\IEEEpeerreviewmaketitle

\section{Introduction}\label{s-introduction}

\IEEEPARstart{T}{he} investigation of multi-agent coordination has regularly attracted contributions from mathematicians, physicists and engineers over the two decades. Classic multi-agent coordination of interest includes consensus \cite{Olfati-Saber07}, flocking \cite{Saber06}, containment control \cite{cao11},  formation control \cite{xiao09}, coverage control \cite{cortes02}, and distributed estimation \cite{yang2008}. Much interest is focusing on dynamic models of agents. Examples include the study of distributed coordination of first-order, second-order, continuous-time and discrete-time dynamic agents. By using the matrix theory, the graph theory, the frequency-domain analysis method and the Lyapunov method etc., lots of results about multi-agent coordination have been obtained \cite{Renbook08,Xiao08-1,ji09,su14}.

As a fundamental problem of distributed coordination, consensus means that a group of agents reach an agreement on the consistent quantity of interest by designing appropriate control input based on local information. Vicsek et al. \cite{vicsek95} proposed a discrete-time model of $n$ agents all moving in the plane with the same speed and demonstrated by simulation that all agents move to one direction. By virtue of graph theory, Jadbabaie et al. \cite{jabdabaie03} explained the consensus behaviour of Vicsek model theoretically and shown that consensus can be achieved if the union of interaction graphs are connected frequently enough. By utilizing the pre-leader-follower decomposition, Wang and Xiao \cite{xiao06-1} studied the state consensus of discrete-time multi-agent systems with bounded time-delays. Second-order consensus of discrete-time multi-agent systems was considered in \cite{lin09} with nonuniform time-delays. Gossip algorithms \cite{boyd06} and broadcast gossip algorithms \cite{aysal09} were also used to analyze the consensus problem, respectively. For continuous-time dynamic agents, Olfati-Saber and Murray \cite{sabertac04} considered the consensus problem of multi-agent systems with switching topologies and time-delays and obtained some necessary and/or sufficient conditions for solving the average consensus. Ren and Beard \cite{ren05} extended the results given in \cite{sabertac04} and presented some more relaxable conditions for solving the consensus. Xie and Wang \cite{Xie07} studied the second-order consensus of multi-agent systems with fixed and switching topologies. Ren \cite{Ren08} investigated the second-order consensus of multi-agent systems in four cases.

Hybrid systems are dynamical systems that involve the interaction of continuous and discrete dynamics. As a class of classic hybrid systems, switched systems have received a major research \cite{Antsaklis2000}. For multi-agent systems, lots of references were concerned with consensus problem under switching topologies \cite{sabertac04,Sun08}. However, Zheng and Wang \cite{Zheng2015} considered the consensus of switched multi-agent system that consists of continuous-time and discrete-time subsystems. They proved that the consensus of switched multi-agent system is solvable under arbitrary switching. Zhu et al. \cite{zhu2015} studied the containment control of such switched multi-agent system. In the practical systems, the dynamics of the agents coupled with each others are different, i.e., the dynamics of agents are hybrid. In general, hybrid means heterogeneous in nature or composition. Different from the previous study, Zheng et al. \cite{zheng11-2} considered the consensus of heterogeneous multi-agent system which is composed of first-order and second-order dynamic agents. The consensus criteria were obtained under different topologies \cite{zheng12-3,zheng12-1}. Finite-time consensus and containment control of the heterogeneous multi-agent system were also studied in \cite{zheng12-2,Zheng14}, respectively.

To the best of our knowledge, however, the existing results of consensus analysis are on the homogeneous and heterogeneous multi-agent systems, i.e., all the agents are continuous-time or discrete-time dynamic behavior at the same time. In the real world, natural and artificial individuals can show collective decision-making. For example, autonomous robots were used to control self-organized behavioral patterns in group-living cockroaches \cite{Halloy2007}. When the continuous-time and discrete-time dynamic agents coexist and interact with each other, it is important to study the consensus problem of such hybrid multi-agent system.  Owing to the coexistence of continuous-time and discrete-time dynamic agents, it is difficult to design the consensus protocol and analyze the consensus problem for hybrid multi-agent system. Up to now, it is not found any approach to analyze the consensus of multi-agent systems with different time-scale. The main objective of this paper is to design the consensus protocols and obtain the consensus criteria of hybrid multi-agent system in different topologies. The main contribution of this paper is threefold. First, we propose the hybrid multi-agent system and give the definition of consensus. Second, three kinds of consensus protocols are presented for hybrid multi-agent system. Finally, by utilizing of the graph theory, some necessary and sufficient conditions are obtained for solving the consensus of hybrid multi-agent system.

The rest of this paper is organized as follows. In Section \ref{s-preliminaries}, we present some notions in graph theory and propose the hybrid multi-agent system. In Section \ref{s-Main results}, three kinds of consensus protocols are provided for solving the consensus of hybrid multi-agent system. In Section \ref{s-Simulations}, numerical simulations are given to illustrate the effectiveness of theoretical results. Finally, some conclusions are drawn in Section \ref{s-Conclusion}.

\textbf{Notation:} Throughout this paper, the following notations will be used: $\mathbb{R}$ denotes the set of real number, $\mathbb{R}^{n}$ denotes the $n-$dimensional real vector space. $\mathcal{I}_{m}=\{1,2, \dots, m\}$, $\mathcal{I}_{n}/\mathcal{I}_{m}=\{m+1,m+2, \dots, n\}$.  For a given vector or matrix $X,$ $X^{T}$ denotes its transpose, $\|X\|$ denotes the Euclidean norm of a vector $X$, $E(X)$ denotes its mathematical expectation. A vector is nonnegative if all its elements are nonnegative. Denote by $\mathbf{1}_n$ (or $\mathbf{0}_n$) the column vector with all entries equal to one (or all zeros). $I_n$ is an $n-$dimensional identity matrix. $\diag\{a_{1},a_{2} \cdots,\ a_{n}\}$ defines a diagonal matrix with diagonal elements being $a_{1}, a_{2} \cdots,\ a_{n}$. Let $\mathbf{e}_i$ denote the canonical vector with a $1$ in the $i$-th entry and 0's elsewhere.

\section{Preliminaries}\label{s-preliminaries}

\subsection{Graph theory}\label{s-graph theory}

A weighted directed graph $\mathscr{G}(\mathscr{A})=(\mathscr{V},\mathscr{E},\mathscr{A})$ of order $n$
consists of a vertex set $\mathscr{V}=\{s_{1}, s_{2}, \cdots, s_{n}\}$,  an edge set
$\mathscr{E}=\{e_{ij}=(s_{i}, s_{j})\}\subset \mathscr{V}\times \mathscr{V}$ and a nonnegative
matrix $\mathscr{A}=[a_{ij}]_{n\times n}$. The neighbor set of the agent $i$ is
$\mathscr{N}_i=\{j: a_{ij}>0\}$. A directed path between two distinct vertices $s_{i}$ and $s_{j}$ is
a finite ordered sequence of distinct edges of $\mathscr{G}$ with the form $(s_{i}, s_{k_{1}}), (s_{k_{1}}, s_{k_{2}}), \cdots, (s_{k_{l}}, s_{j})$.
A directed tree is a directed graph, where there exists a vertex called the root such that there exists a unique directed path from this vertex to every other vertex. A directed spanning tree is a directed tree, which consists of all the nodes and some edges in $\mathscr{G}$.
If a directed graph has the property that $(s_{i}, s_{j})\in \mathscr{E} \Leftrightarrow (s_{j}, s_{i})\in \mathscr{E}$, the directed graph is called undirected. An undirected graph is said to be connected if there exists a path between any two distinct vertices of the graph. The degree matrix
$\mathscr{D}=[d_{ij}]_{n\times n}$ is a diagonal matrix with $d_{ii}=\sum_{j:s_j\in \mathscr{N}_{i}} a_{ij}$ and
the Laplacian matrix of the graph is defined as $\mathscr{L}=[l_{ij}]_{n \times n}=\mathscr{D}-\mathscr{A}.$
It is easy to see that $\mathscr{L}\mathbf{1}_{n}=0$. 

A nonnegative matrix is said to be a (row) stochastic matrix if all its row sums are $1$. A stochastic matrix $P=[p_{ij}]_{n\times n}$ is called indecomposable and aperiodic (SIA) if $\lim_{k\rightarrow\infty} P^k = \mathbf{1}y^T$ , where $y$ is some column vector. $\mathscr{G}$ is said the graph associated with $P$ when $(s_{i}, s_{j})\in \mathscr{E}$ if and only if $p_{ji}>0$. The following results propose the relationship between a stochastic matrix and its associated graph.

\begin{lemma}\label{lemma-renw-3.5}(\cite{ren05})  
 A stochastic matrix has algebraic multiplicity equal
to one for its eigenvalue $\lambda = 1$ if and only if the graph associated
with the matrix has a spanning tree. Furthermore, a stochastic matrix
with positive diagonal elements has the property that $|\lambda| < 1$ for every
eigenvalue not equal to one.
\end{lemma}

\begin{lemma}\label{lemma-renw-3.7}(\cite{ren05}) 
 Let $A = [a_{ij} ]_{n\times n}$ be a stochastic matrix. If
A has an eigenvalue $\lambda = 1$ with algebraic multiplicity equal to one,
and all the other eigenvalues satisfy $|\lambda| < 1$, then $A$ is SIA, that is,
$\lim_{m\rightarrow \infty} A^m=\mathbf{1}_n\nu^T $, where $\nu$ satisfies $A^T\nu = \nu$ and $\mathbf{1}_n^T \nu= 1$. Furthermore, each element of $\nu$ is nonnegative.
\end{lemma}

Based on Lemma \ref{lemma-renw-3.5} and Lemma \ref{lemma-renw-3.7}, we give the following result which will be used in the next content.

\begin{lemma}\label{lemma-1}
Let $H=\diag\{h_1,h_2,\dots,h_n\}$ and $0<h_i<\frac{1}{d_{ii}}$, $i\in \mathcal{I}_n$. Then, $I_n-H\mathscr{L}$ is SIA, i.e., $\lim_{k\rightarrow \infty}\left[I_n-H\mathscr{L}\right]^k=\mathbf{1}_n\nu^T$, if and only if graph $\mathscr{G}$ has a spanning tree. Furthermore, $\left[I_n-H\mathscr{L}\right]^T\nu  = \nu$, $\mathbf{1}_n^T \nu = 1$ and each element of $\nu$ is nonnegative.
\end{lemma}
{\it Proof.} \textbf{(Sufficiency)} Let $P=I_n-H\mathscr{L}$. From the definition of $\mathscr{L}$, we have $P=K+H\mathscr{A}$ where $K=I_n-H\mathscr{D}$. It follows from $h_i\in \left(0,\frac{1}{d_{ii}}\right)$ that $K$ is a positive diagonal matrix. Consequently, it is easy to obtain that $P$ is a stochastic matrix with positive diagonal entries. Obviously, for all $i,j\in \mathcal{I}_n,~i\neq j$, the $(i,j)$-th entry of $P$ is positive if and only if $a_{ij}>0$. Then, $\mathscr{G}$ is the graph associated with $P$. Combining Lemma \ref{lemma-renw-3.5} and Lemma \ref{lemma-renw-3.7}, we have when graph $\mathscr{G}$ has a spanning tree, $\lim_{k\rightarrow \infty}P^k=\mathbf{1}_n\nu^k$ where $\nu$ is a nonnegative vector. Moreover, $\nu$ satisfies $P^T\nu  = \nu$ and $\mathbf{1}_n^T \nu =1$.

\textbf{(Necessary)} From Lemma \ref{lemma-renw-3.5}, if $\mathscr{G}$ does not have a spanning tree, the algebraic multiplicity of eigenvalue $\lambda= 1$ of $P$ is $m>1$. Then, it is easy to prove that the rank of $\lim_{k\rightarrow \infty}P^k$ is greater than 1, which implies that $\lim_{k\rightarrow \infty}P^k\neq \mathbf{1}_n\nu^k$. $\blacksquare$

\subsection{Hybrid multi-agent system}\label{s-Hybrid multi-agent systems}

Suppose that the hybrid multi-agent system consists of continuous-time and discrete-time dynamic agents. The number of agents is $n$, labelled $1$ through $n$, where the number of continuous-time dynamic agents is $m$ $(m<n)$. Without loss of generality, we assume that agent $1$ through agent $m$ are continuous-time dynamic agents. Then, each agent has the dynamics as follows:
\begin{equation}\label{mod-hybrid}
\left\{
   \begin{aligned}
   &\dot{x}_{i}(t)=u_{i}(t), &&i \in \mathcal{I}_{m}, \\
   &x_{i}(t_{k+1})=x_{i}(t_{k})+u_{i}(t_{k}), t_k=kh,~~k\in \mathbb{N}, &&i\in\mathcal{I}_{n}/\mathcal{I}_{m},
   \end{aligned}
   \right.
  \end{equation}
where $h$ is the sampling period, $x_{i}\in \mathbb{R}$ and $u_{i}\in \mathbb{R}$ are the position-like and control input of agent $i$, respectively. The initial conditions are $x_{i}(0)=x_{i0}$.  Let $x(0)=[x_{10}, x_{20}, \cdots, x_{n0}]^{T}$.

Each agent is regraded as a vertex in a graph. Each edge $(s_{i}, s_{j})\in \mathscr{E}$ corresponds to an available information link from agent $i$ to agent $j$. Moreover, each agent updates its current state based on the information received from its neighbors. In this paper, we suppose that there exists communication behavior in hybrid multi-agent system (\ref{mod-hybrid}), i.e.,  there are agent $i$ and agent $j$ which make $a_{ij}>0$.

\begin{definition}\label{def-consensus} Hybrid multi-agent system (\ref{mod-hybrid}) is said to reach consensus if for any initial conditions, we have
\begin{equation}\label{lim-tk-1}
\lim_{t_{k} \to \infty }\|x_{i}(t_{k})-x_{j}(t_{k})\|=0, ~~for~~i,j\in \mathcal{I}_{n},
\end{equation}
and
\begin{equation}\label{lim-t-1}
   \lim_{t \to \infty }\|x_{i}(t)-x_{j}(t)\|=0, ~~for~~i,j\in \mathcal{I}_{m}.
\end{equation}
\end{definition}


\section{Main results}\label{s-Main results}

In this section, the consensus problem of hybrid multi-agent system (\ref{mod-hybrid}) will be considered under three kinds of control inputs (consensus protocols) respectively.

\subsection{Case 1}\label{s-case1}
In this subsection, we assume that all agents communicate with their neighbours and update their control inputs in the sampling time $t_k$. Then, the consensus protocol for hybrid multi-agent system (\ref{mod-hybrid}) is presented as follows.
\begin{equation}\label{control-input-1}
\left\{
\begin{aligned}
&u_i(t)=\sum_{j=1}^na_{ij}(x_j(t_k)-x_i(t_k)),~~t\in (t_k,t_{k+1}],~~&&i\in \mathcal{I}_m,\\
&u_i(t_k)=h\sum_{j=1}^na_{ij}(x_j(t_k)-x_i(t_k)), &&i\in\mathcal{I}_{n}/\mathcal{I}_{m},
\end{aligned}
\right.
\end{equation}
where $\mathscr{A}=[a_{ij}]_{n\times n}$ is the weighted adjacency matrix
associated with the graph $\mathscr{G}$, $h=t_{k+1}-t_{k}>0$ is the sampling period.

\begin{theorem}\label{consensus-case1}
Consider a directed communication network $\mathscr{G}$ and suppose that $h<\frac{1}{\max_{i\in \mathcal{I}_n}\{d_{ii}\}}$. Then, hybrid multi-agent system (\ref{mod-hybrid}) with protocol (\ref{control-input-1}) can solve consensus if and only if the network $\mathscr{G}$ has a directed spanning tree. Moreover, the consensus state is $\nu^{T}_{1}x(0)$, where $\mathscr{L}^T\nu_{1}=0$.  
\end{theorem}

{\it Proof.} \textbf{(Sufficiency)} Firstly, we will prove that equation (\ref{lim-tk-1}) holds. From (\ref{control-input-1}), we know that
\begin{equation}\label{state-t-1}
\left\{
\begin{aligned}
&x_i(t)=x_i(t_k)+(t-t_k)\sum_{j=1}^na_{ij}(x_j(t_k)-x_i(t_k)),~~t\in (t_k,t_{k+1}], ~~&&i\in \mathcal{I}_m,\\
&x_i(t_{k+1})=x_i(t_k)+h\sum_{j=1}^na_{ij}(x_j(t_k)-x_i(t_k)), && i\in\mathcal{I}_{n}/\mathcal{I}_{m}.
\end{aligned}
\right.
\end{equation}
Therefore, it follows that
\begin{equation}\label{state-tk-1}
x_i(t_{k+1})=x_i(t_k)+h\sum_{j=1}^na_{ij}(x_j(t_k)-x_i(t_k)), ~~ i\in \mathcal{I}_n.
\end{equation}

Let $x(t_k)=[x_1(t_k),x_2(t_k),...,x_n(t_k)]^T$. Then, equation (\ref{state-tk-1}) can be written in matrix form as
\begin{equation}\label{state-mtr-1}
x(t_{k+1})=(I_n-h\mathscr{L})x(t_k).
\end{equation}
According to Lemma \ref{lemma-1}, since $\mathscr{G}$ has a directed spanning tree and $h<\frac{1}{\max_{i\in \mathcal{I}_n}\{d_{ii}\}}$, we have
$\lim_{k\rightarrow \infty} (I_n-h\mathscr{L})^k=\mathbf{1}_n\nu^{T}_{1}$,
where $(I_n-h\mathscr{L})^T\nu_{1}=\nu_{1}$. Thus, it is easy to obtain
\[
\lim_{t_k\rightarrow \infty}x(t_{k})=\lim_{k\rightarrow \infty}(I_n-h\mathscr{L})^kx(0)=\mathbf{1}_n\nu^{T}_{1} x(0)
\]
and $\mathscr{L}^T\nu_{1}=0$.
As a consequence, equation (\ref{lim-tk-1}) holds. Moreover,
\[
\lim_{t_k\rightarrow \infty}x_i(t_{k})=\nu^{T}_{1} x(0),~~for~~i\in \mathcal{I}_{n}.
\]

Secondly, we have
\[
\|x_i(t)-x_j(t)\|\leq \|x_i(t)-x_i(t_k)\|+\|x_i(t_{k})-x_j(t_k)\|+\|x_j(t_{k})-x_j(t)\|.
\]
From (\ref{state-t-1}),  it is easy to know that
\[
\|x_i(t)-x_i(t_k)\|\leq h\sum_{j=1}^na_{ij}\|x_j(t_k)-x_i(t_k)\|, ~~for~~t\in (t_k,t_{k+1}].
\]
when $t \rightarrow \infty$, we have $t_k \rightarrow \infty$. Thus,
$$
\lim_{t\rightarrow \infty}\|x_i(t)-x_i(t_k)\|=0,~~i\in\mathcal{I}_{m}.
$$
Moreover,
$$\lim_{t\rightarrow \infty}x_i(t)=\lim_{t_k\rightarrow \infty}x_i(t_{k})=\nu^T x(0),~~i\in\mathcal{I}_{m},$$
which implies that equation (\ref{lim-t-1}) holds.

Therefore, hybrid multi-agent system (\ref{mod-hybrid}) with protocol (\ref{control-input-1}) can solve consensus probem with consensus state $\nu^{T}_{1}x(0)$, where $\mathscr{L}^T\nu_{1}=0$.

\textbf{(Necessity)} From Lemma \ref{lemma-1}, we have $\lim_{k\rightarrow \infty} (I_n-h\mathscr{L})^k\neq \mathbf{1}_n\nu^{T}_{1}$ when $\mathscr{G}$ does not have a directed spanning tree. Therefore, equation (\ref{lim-tk-1}) does not hold, which implies that hybrid multi-agent system (\ref{mod-hybrid}) can not reach consensus. $\blacksquare$



\subsection{Case 2}\label{s-case2}

In this subsection, we still assume that the interaction among agents happens in sampling time $t_k$. However, different from Case 1, we assume that each continuous-time dynamic agent can observe its own state in real time. Thus, the consensus protocol for hybrid multi-agent system (\ref{mod-hybrid}) is presented as follows. 
\begin{equation}\label{control-input-2}
\left\{
\begin{aligned}
&u_i(t)=\sum_{j=1}^na_{ij}(x_j(t_k)-x_i(t)),~~t\in (t_k,t_{k+1}], &&i\in \mathcal{I}_m,\\
&u_i(t_k)=h\sum_{j=1}^na_{ij}(x_j(t_k)-x_i(t_k)), &&i\in\mathcal{I}_{n}/\mathcal{I}_{m},
\end{aligned}
\right.
\end{equation}
where $\mathscr{A}=[a_{ij}]_{n\times n}$ is the weighted adjacency matrix
associated with the graph $\mathscr{G}$, $h=t_{k+1}-t_{k}>0$ is the sampling period.

\begin{theorem}\label{consensus-case2}
Consider a directed communication network $\mathscr{G}$ and suppose that $h<\frac{1}{\max_{i\in\mathcal{I}_{n}/\mathcal{I}_{m}}\{d_{ii}\}}$. Then, hybrid multi-agent system (\ref{mod-hybrid}) with protocol (\ref{control-input-2}) can solve consensus if and only if the network $\mathscr{G}$ has a directed spanning tree. Moreover, the consensus state is $\nu^{T}x(0)$, where $\mathscr{L}^TH\nu=0$,  $H=\diag\left\{\frac{1-e^{-\sum_{j=1}^na_{1j}h}}{\sum_{j=1}^na_{1j}},
\dots,\frac{1-e^{-\sum_{j=1}^na_{mj}h}}{\sum_{j=1}^na_{mj}},h,\dots,h\right\}$.
\end{theorem}

{\it Proof.} \textbf{(Sufficiency)} Firstly, we will prove that equation (\ref{lim-tk-1}) holds. Solving hybrid multi-agent system (\ref{mod-hybrid}) with protocol (\ref{control-input-2}), we have
\begin{equation}\label{state-t-2}
\left\{
\begin{aligned}
&x_i(t)=x_i(t_k)+\frac{1-e^{-\sum_{j=1}^na_{ij}(t-t_k)}}{\sum_{j=1}^na_{ij}}\sum_{j=1}^na_{ij}(x_j(t_k)-x_i(t_k)),~t\in (t_k,t_{k+1}], &&i\in \mathcal{I}_m,\\
&x_i(t_{k+1})=x_i(t_k)+h\sum_{j=1}^na_{ij}(x_j(t_k)-x_i(t_k)), && i\in\mathcal{I}_{n}/\mathcal{I}_{m}.
\end{aligned}
\right.
\end{equation}
Accordingly, at time $t_{k+1}$, the states of agents are
\begin{equation}\label{state-tk-2}
\left\{
\begin{aligned}
&x_i(t_{k+1})=x_i(t_k)+\frac{1-e^{-\sum_{j=1}^na_{ij}h}}{\sum_{j=1}^na_{ij}}\sum_{j=1}^na_{ij}(x_j(t_k)-x_i(t_k)), ~~&&i\in \mathcal{I}_m,\\
&x_i(t_{k+1})=x_i(t_k)+h\sum_{j=1}^na_{ij}(x_j(t_k)-x_i(t_k)), && i\in\mathcal{I}_{n}/\mathcal{I}_{m}.
\end{aligned}
\right.
\end{equation}
Hence, (\ref{state-tk-2}) can be rewritten in compact form as
\begin{equation}\label{state-tk-2m}
x(t_{k+1})=(I_n-H\mathscr{L})x(t_{k}),
\end{equation}
where $x(t_k)=[x_1(t_k),x_2(t_k),...,x_n(t_k)]^T$, $H=\diag\left\{\frac{1-e^{-\sum_{j=1}^na_{1j}h}}{\sum_{j=1}^na_{1j}},
\dots,\frac{1-e^{-\sum_{j=1}^na_{mj}h}}{\sum_{j=1}^na_{mj}},h,\dots,h\right\}$.

It is easy to know that $\frac{1-e^{-\sum_{j=1}^na_{ij}h}}{\sum_{j=1}^na_{ij}}<\frac{1}{d_{ii}}$, $i\in \mathcal{I}_m$. Owing to $h<\frac{1}{\max_{i\in\mathcal{I}_{n}/\mathcal{I}_{m}}\{d_{ii}\}}$, we have $0<h_{i}<\frac{1}{d_{ii}}$ for $H$.  From Lemma \ref{lemma-1}, because $\mathscr{G}$ has a directed spanning tree, $I_n-H\mathscr{L}$ is an SIA matrix, i.e.,
$\lim_{k\rightarrow \infty} (I_n-H\mathscr{L})^k=\mathbf{1}_n\nu^T$,
where $(I_n-H\mathscr{L})^T\nu=\nu$. Hence,
\[
\lim_{t_k\rightarrow \infty}x(t_{k})=\lim_{k\rightarrow \infty}(I_n-H\mathscr{L})^kx(0)=\mathbf{1}_n\nu^T x(0),
\]
which means that
\[
\lim_{t_k\rightarrow \infty}x_i(t_{k})=\nu^T x(0),~~for~~i\in \mathcal{I}_{n}.
\]
Obviously, equation (\ref{lim-tk-1}) holds. Moreover, it follows from $(I_n-H\mathscr{L})^T\nu=\nu$ that $\mathscr{L}^TH\nu=0$.

Secondly, we know from (\ref{state-t-2}) that
\[
\|x_i(t)-x_i(t_k)\|\leq\frac{1}{\sum_{i=1}^na_{ij}}\sum_{j=1}^na_{ij}\|x_j(t_k)-x_i(t_k)\|, ~~t\in (t_k,t_{k+1}],~~i\in \mathcal{I}_m.
\]
Since $t_k\rightarrow \infty$ when $t\rightarrow \infty$,
\[
\lim_{t\rightarrow \infty}\|x_i(t)-x_i(t_k)\|=0.
\]
Thus, we have
$\lim_{t\rightarrow \infty}x_i(t)=\nu^T x(0),~~i\in\mathcal{I}_{m}.$
Consequently, equation (\ref{lim-t-1}) holds.

Therefore, from Definition \ref{def-consensus}, hybrid multi-agent system (\ref{mod-hybrid}) with protocol (\ref{control-input-2}) reaches consensus. Moreover, the consensus state is $\nu^T x(0)$.

\textbf{(Necessity)} Similar to the proof of necessity in Theorem \ref{consensus-case1}, we know that if the directed communication network $\mathscr{G}$ does not have a directed spanning tree, then hybrid multi-agent system (\ref{mod-hybrid}) can not achieve consensus. $\blacksquare$

\begin{remark}\label{h}
Compared with the restriction of sampling period $h$ in Theorem \ref{consensus-case1} and Theorem \ref{consensus-case2}, it is easy to find that $h$ is only related to the out-degrees of discrete-time dynamic agents if each continuous-time dynamic agent can observe the real time information.
\end{remark}

\subsection{Case 3}\label{s-case3}

In this subsection, we assume that all agents interact with each other in a gossip-like manner. Some assumptions are given as follows.

(\textbf{A1}) The communication network of hybrid multi-agent system (\ref{mod-hybrid}) is undirected, i.e., $a_{ij}=a_{ji}$ for all $i,j\in \mathcal{I}_n$.

(\textbf{A2}) At time $t_k$, two agents $i$ and $j$ ($i<j$) satisfying $(s_i,s_j)\in \mathscr{E}$ are chosen with probability $p_{ij}$, where $p_{ij}\in (0,1)$ and $\sum_{(s_i,s_j)\in \mathscr{E}} p_{ij}=1$.

When agents $i$ and $j$ are chosen, their interplay follows the below situations, where $h=t_{k+1}-t_{k}>0$ is the sampling period.
\begin{itemize}
  \item If $i$ and $j$ are continuous-time dynamic agents, i.e., $i,j \in \mathcal{I}_m$, they will communicate during $(t_k,t_{k+1}]$. The control inputs of two agents are
      \begin{equation}\label{gossip-CC}
      \left\{
      \begin{aligned}
         &u_i(t)=a_{ij}(x_j(t)-x_i(t)),~~&&t\in(t_k,t_{k+1}],~~i\in \mathcal{I}_m,\\
         &u_j(t)=a_{ji}(x_i(t)-x_j(t)),~~&&t\in(t_k,t_{k+1}],~~j\in \mathcal{I}_m.
      \end{aligned}
      \right.
      \end{equation}
  \item If $i$ is continuous-time dynamic agent and $j$ is discrete-time dynamic agent, i.e., $i\in \mathcal{I}_m,~j\in \mathcal{I}_{n}/\mathcal{I}_{m}$, the control inputs of two agents are
      \begin{equation}\label{gossip-CD}
        \left\{
       \begin{aligned}
      &u_i(t)=a_{ij}(x_j(t_{k})-x_i(t)),~&&i\in \mathcal{I}_m,\\
      &u_j(t_{k+1})=x_j(t_k)+ha_{ji}(x_i(t_k)-x_j(t_k)),~&&j\in \mathcal{I}_{n}/\mathcal{I}_{m}.
     \end{aligned}
      \right.
      \end{equation}
  \item If $i$ and $j$ are discrete-time dynamic agents, i.e., $i,j \in \mathcal{I}_{n}/\mathcal{I}_{m}$, the control inputs of two agents are
      \begin{equation}\label{gossip-DD}
        \left\{
      \begin{aligned}
        &u_i(t_{k+1})=x_i(t_k)+ha_{ij}(x_j(t)-x_i(t_k)),~&&i\in \mathcal{I}_{n}/\mathcal{I}_{m},\\
        &u_j(t_{k+1})=x_j(t_k)+ha_{ji}(x_i(t_k)-x_j(t_k)),~&&j\in \mathcal{I}_{n}/\mathcal{I}_{m}.
      \end{aligned}
        \right.
      \end{equation}
\end{itemize}
For each $r\in\mathcal{I}_n/\{i,j\}$, it keeps static, i.e.,
\begin{equation}\label{gossip-lf}
\left\{
\begin{aligned}
&u_r(t)=0,~&&r\in\mathcal{I}_m,\\
&u_r(t_{k})=x_r(t_k),~&&r\in \mathcal{I}_{n}/\mathcal{I}_{m}.
\end{aligned}
\right.
\end{equation}


\begin{theorem}\label{consensus-gossip}
Consider an undirected communication network $\mathscr{G}$. Assume that (A1)--(A2) hold and $h<\frac{1}{\max_{i,j\in \mathcal{I}_n}\{a_{ij}\}}$. Then, hybrid multi-agent system (\ref{mod-hybrid}) with control input (\ref{gossip-CC})--(\ref{gossip-lf}) can solve consensus in mean sense if and only if the network $\mathscr{G}$ is connected.
\end{theorem}
{\it Proof.} \textbf{(Sufficiency)} It suffices to prove that
\begin{equation}\label{E-lim-tk-1}
\lim_{t_{k} \to \infty }\mathbb{E}\|x_{i}(t_{k})-x_{j}(t_{k})\|=0, ~~for~~i,j\in \mathcal{I}_{n},
\end{equation}
and
\begin{equation}\label{E-lim-t-1}
   \lim_{t \to \infty }\mathbb{E}\|x_{i}(t)-x_{j}(t)\|=0, ~~for~~i,j\in \mathcal{I}_{m},
\end{equation}
hold for any initial states.

Firstly, we will show that equation (\ref{E-lim-tk-1}) holds. From (\ref{gossip-CC})--(\ref{gossip-lf}), if agents $i$ and $j$ are selected to interplay at time $t_{k}$, the states of all agents at time $t_{k+1}$ are three cases:
\begin{subequations}\label{gossip-states}
\begin{align}
&(I)~~&&
\left\{
\begin{aligned}
&x_i(t_{k+1})=x_i(t_{k})+\frac{1-e^{-2a_{ij}h}}{2}(x_j(t_{k})-x_i(t_{k})),~&&i\in \mathcal{I}_m,\\
&x_j(t_{k+1})=x_j(t_{k})+\frac{1-e^{-2a_{ij}h}}{2}(x_i(t_{k})-x_j(t_{k})),~&&j\in \mathcal{I}_m,\\
&x_r(t_{k+1})=x_r(t_{k}),~~&&r\in\mathcal{I}_n/\{i,j\},
\end{aligned}
\right. \label{GS-CC} \\
&(II)~~&&
\left\{
\begin{aligned}
&x_i(t_{k+1})=x_i(t_k)+(1-e^{-a_{ij}h})(x_j(t_k)-x_i(t_k)),~&&i\in \mathcal{I}_m,\\
&x_j(t_{k+1})=x_j(t_{k})+ha_{ji}(x_i(t_k)-x_j(t_k)),&&j\in \mathcal{I}_{n}/\mathcal{I}_{m},\\
&x_r(t_{k+1})=x_r(t_{k}),~~&&r\in\mathcal{I}_n/\{i,j\},
\end{aligned}
\right.\label{GS-CD}\\
&(III)~~&&
\left\{
\begin{aligned}
&x_i(t_{k+1})=x_i(t_k)+ha_{ij}(x_j(t)-x_i(t_k)),~&&j\in \mathcal{I}_{n}/\mathcal{I}_{m},\\
&x_j(t_{k+1})=x_j(t_k)+ha_{ji}(x_i(t_k)-x_j(t_k)),~&&j\in \mathcal{I}_{n}/\mathcal{I}_{m},\\
&x_r(t_{k+1})=x_r(t_{k}),~~&&r\in\mathcal{I}_n/\{i,j\}.
\end{aligned}
\right.\label{GS-DD}
\end{align}
\end{subequations}
Note that (\ref{gossip-states}) can be rewritten in the following compact form as
$$x(t_{k+1})=\Phi_{ij}x(t_k),$$
where
\begin{equation}\label{phi-ij}
\left\{
\begin{aligned}
&\Phi_{ij}=I_n-\frac{1-e^{-2a_{ij}h}}{2}(\mathbf{e}_i-\mathbf{e}_j)(\mathbf{e}_i-\mathbf{e}_j)^T,&&i,j\in \mathcal{I}_m,\\
&\Phi_{ij}=I_n-(1-e^{-a_{ij}h})\mathbf{e}_i(\mathbf{e}_i-\mathbf{e}_j)^T
-ha_{ij}\mathbf{e}_j(\mathbf{e}_j-\mathbf{e}_i)^T,&&i\in \mathcal{I}_m,j\in \mathcal{I}_{n}/\mathcal{I}_{m},\\
&\Phi_{ij}=I_n-ha_{ij}(\mathbf{e}_i-\mathbf{e}_j)(\mathbf{e}_i-\mathbf{e}_j)^T,&&i,j\in \mathcal{I}_{n}/\mathcal{I}_{m}.
\end{aligned}
\right.
\end{equation}
According to (A2), it is not hard to know that $\{x({t_k})\}$ is a stochastic
linear system
\[
x(t_{k+1})=\Phi_{k}x(t_k), ~~\mathbb{P}(\Phi_{k} =\Phi_{ij})\xlongequal [ ]{i.i.d.}p_{ij}.
\]
Therefore, it follows that
\begin{equation}\label{gossip-Estate-tk}
\mathbb{E}(x(t_{k+1}))=\mathbb{E}(\Phi_{k})\mathbb{E}(x(t_k)).
\end{equation}

Due to $0<h<\frac{1}{\max_{i,j\in \mathcal{I}_n}\{a_{ij}\}}$, we have $\frac{1-e^{-2a_{ij}h}}{2}\in (0,\frac{1}{2})$, $1-e^{-a_{ij}h}\in (0,1)$ and $ha_{ij}\in (0,1)$. Thus, $\Phi_{ij}$ is a stochastic matrix with positive diagonal entries. Moreover, the $(i,j)$-th and $(j,i)$-th entries of $\Phi_{ij}$ are positive, while all other non-diagonal entries are zeros. Hence, noticing that
$$\mathbb{E}(\Phi_{k})=\sum_{(s_i,s_j)\in \mathscr{E}}\Phi_{ij}p_{ij},$$
we know that $\mathbb{E}(\Phi_{k})$ is a stochastic matrix satisfying
\begin{itemize}
  \item all diagonal entries are positive;
  \item the $(i,j)$-th and $(j,i)$-th entries are positive if and only if $(s_i,s_j)\in \mathscr{E}$.
\end{itemize}

Consequently, $\mathscr{G}$ is the graph associated with $\mathbb{E}(\Phi_{k})$. Since $\mathscr{G}$ is connected, combining Lemma \ref{lemma-renw-3.5} and Lemma \ref{lemma-renw-3.7}, we have
$$\lim_{k\rightarrow\infty}(\mathbb{E}(\Phi_{k}))^k=\mathbf{1}_n\nu^{'}.$$
From (\ref{gossip-Estate-tk}), we have
$$
\lim_{t_k\rightarrow\infty}\mathbb{E}(x(t_{k}))
=\lim_{k\rightarrow\infty}(\mathbb{E}(\Phi_{k}))^kx(0)=\mathbf{1}_n\nu^{'}x(0),
$$
which implies that equation (\ref{E-lim-tk-1}) holds.

Secondly, at time $t\in(t_k,t_{k+1}]$, the state of each $i\in \mathcal{I}_m$ follows the below three scenarios:
\begin{itemize}
  \item if~$i$~is~selected~to~communicate~with~its~ continuous-time dynamical~neighbour~$j\in \mathcal{I}_m$, $$x_i(t)=\frac{1+e^{-2a_{ij}(t-t_k)}}{2}x_i(t_{k})+\frac{1-e^{-2a_{ij}(t-t_k)}}{2}x_j(t_{k});$$
  \item if~$i$~is~selected~to~communicate~with~its~ discrete-time dynamical~neighbour~$j\in \mathcal{I}_{n}/\mathcal{I}_{m}$,
  $$x_i(t)=e^{-a_{ij}(t-t_k)}x_i(t_k)+(1-e^{-a_{ij}(t-t_k)})x_j(t_k);$$
  \item if~$i$ is not selected, $x_i(t)=x_i(t_k).$
\end{itemize}
Thus, we have
\begin{equation}\label{gossip-Estate-t}
\begin{aligned}
 \mathbb{E}(x_i(t))&=\left(1-\sum_{j\in \mathscr{N}_{i}}p_{ij}\right) \mathbb{E}(x_i(t_k))\\
 &+\sum_{j\in \mathscr{N}_{i}}p_{ij}\left[\beta_{ij} \mathbb{E}(x_i(t_k))+(1-\beta_{ij}) \mathbb{E}(x_j(t_k))\right],~~t\in(t_k,t_{k+1}],
 \end{aligned}
\end{equation}
where
$$
\beta_{ij}=\left\{
\begin{aligned}
&\frac{1+e^{-2a_{ij}(t-t_k)}}{2},~&&j\in \mathcal{I}_m\\
&e^{-a_{ij}(t-t_k)},~&&j\in \mathcal{I}_{n}/\mathcal{I}_{m}.
\end{aligned}
\right.
$$
Since $t_k\rightarrow \infty$ when $t\rightarrow \infty$, it is easy to obtain from (\ref{gossip-Estate-t}) that
$$\lim_{t\rightarrow\infty}\mathbb{E}(x(t))=\lim_{t_k\rightarrow\infty}\mathbb{E}(x(t_{k}))=\mathbf{1}_nv^{'}x(0)$$
holds for all $i\in \mathcal{I}_m$, which means equation (\ref{E-lim-t-1}) holds.

Therefore, hybrid multi-agent system (\ref{mod-hybrid}) with control inputs (\ref{gossip-CC})--(\ref{gossip-lf}) can solve consensus problem in mean sense.

\textbf{(Necessity)} When $\mathscr{G}$ is not connected, similar to the proof of necessity in Theorem \ref{consensus-case1}, we know that hybrid multi-agent system (\ref{mod-hybrid}) can not reach consensus. $\blacksquare$

\begin{remark}\label{unified}
Note that hybrid multi-agent system (\ref{mod-hybrid}) presents a unified framework for both the discrete-time multi-agent system and the continuous-time multi-agent system. In other words, if $m=0$, hybrid multi-agent system (\ref{mod-hybrid}) becomes a discrete-time multi-agent system. And if $m=n$, hybrid multi-agent system (\ref{mod-hybrid}) becomes a continuous-time multi-agent system.
\end{remark}
\section{Simulations }\label{s-Simulations}

In this section, we provided some simulations to demonstrate the effectiveness of the theoretical results in this paper.

Suppose that there are 6 agents. The continuous-time dynamic agents and the discrete-time dynamic agents are denoted by 1--3 and 4--6, respectively.
In the following, all networks with 0-1 weights will be needed. Let $x(0)=[-13,14,3,-9,-3,6]^T$ and $h=0.2$.

\begin{example}\label{ex1}
\begin{figure}[htbp]
  \centering
  \includegraphics[scale=0.3]{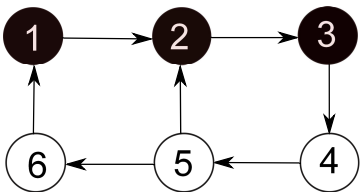}\\
  \caption{A directed graph $\mathscr{G}_1$.}\label{graph-ex1}
\end{figure}

The communication network $\mathscr{G}_{1}$ is shown in Fig. \ref{graph-ex1}. It can be noted that $\mathscr{G}_{1}$ has a directed spanning tree. It is easy to calculate that the sampling period $h<\frac{1}{\max_{i=1}^6\{d_{ii}\}}$. By using consensus protocol (\ref{control-input-1}), the state trajectories of all the agents are shown in Fig. \ref{result-ex1}, which is consistent with the sufficiency of Theorem \ref{consensus-case1}.

\begin{figure}[htbp]
  \centering
  \includegraphics[width=10.00cm]{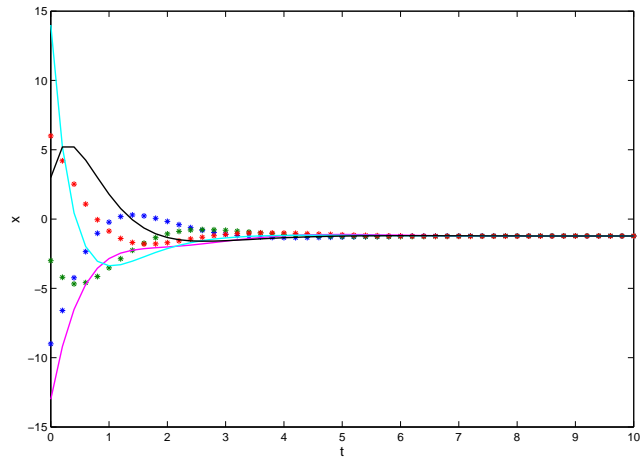}\\
  \caption{The state trajectories of all the agents with consensus protocol (\ref{control-input-1}) and communication network $\mathscr{G}_1$.}\label{result-ex1}
\end{figure}

\end{example}

\begin{example}\label{ex2}

\begin{figure}[htbp]
  \centering
  \includegraphics[scale=0.3]{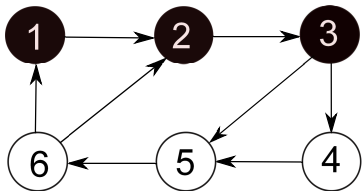}\\
  \caption{A directed graph $\mathscr{G}_2$.}\label{graph-ex2}
\end{figure}

Assume that the communication network $\mathscr{G}_2$ is shown in Fig. \ref{graph-ex2}. It is easy to know that $\mathscr{G}_2$ has a directed spanning tree. By calculation, the sampling period $h<\frac{1}{\max_{i=4}^6\{d_{ii}\}}$. By using consensus protocol (\ref{control-input-2}), the state trajectories of all the agents are shown in Fig. \ref{result-ex2}, which is consistent with the results in Theorem \ref{consensus-case2}.

\begin{figure}[htbp]
  \centering
  \includegraphics[width=10.00cm]{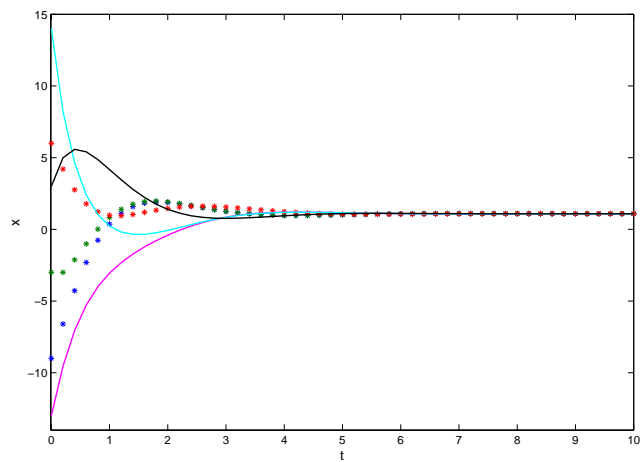}\\
  \caption{The state trajectories of all the agents with consensus protocol (\ref{control-input-2}) and communication network $\mathscr{G}_2$.}\label{result-ex2}
\end{figure}

\end{example}

\begin{figure}[htbp]
  \centering
  \includegraphics[scale=0.3]{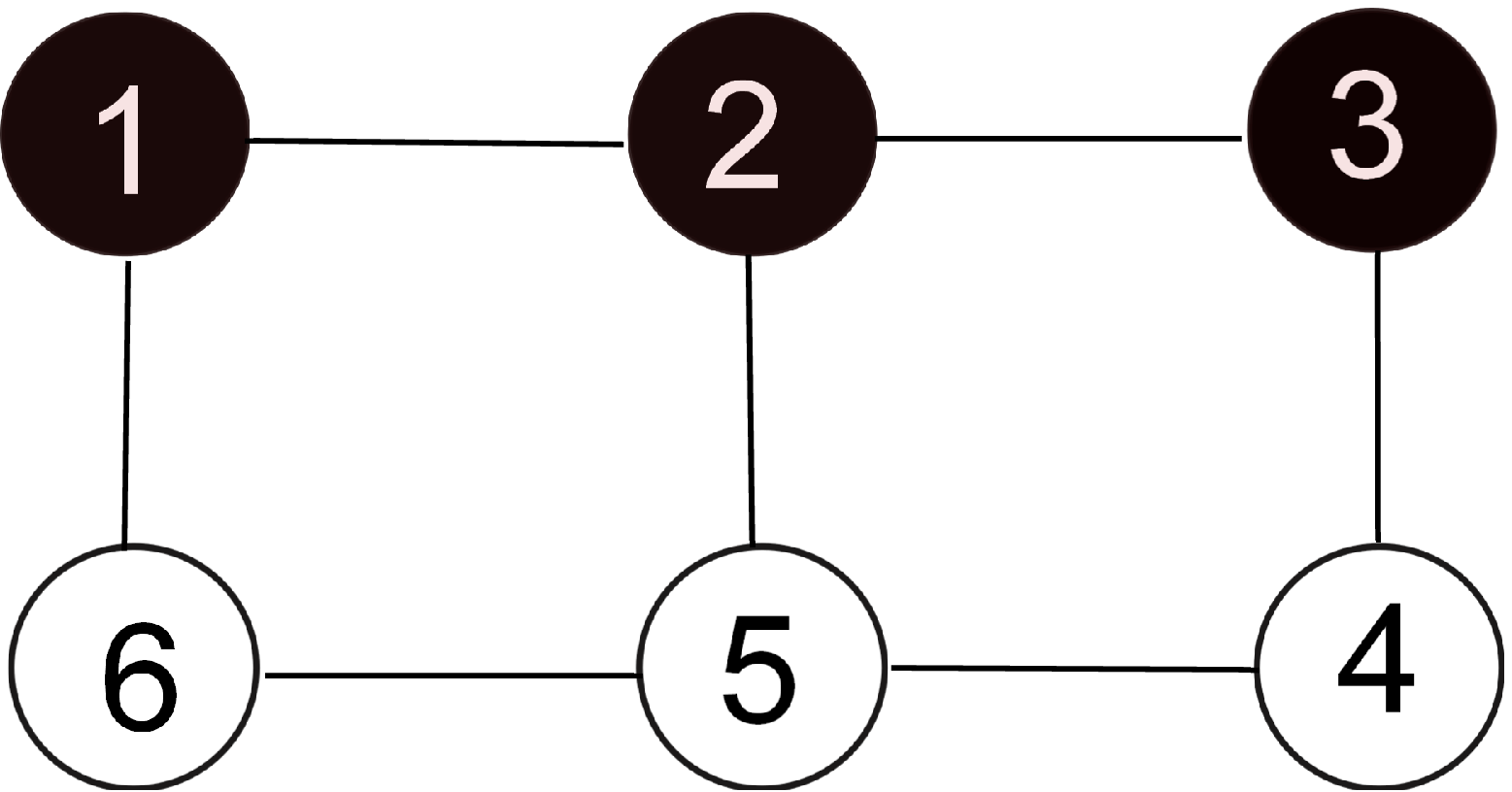}\\
  \caption{An undirected graph $\mathscr{G}_3$.}\label{graph-ex3}
\end{figure}

\begin{figure}[htbp]
  \centering
  \includegraphics[width=10.00cm]{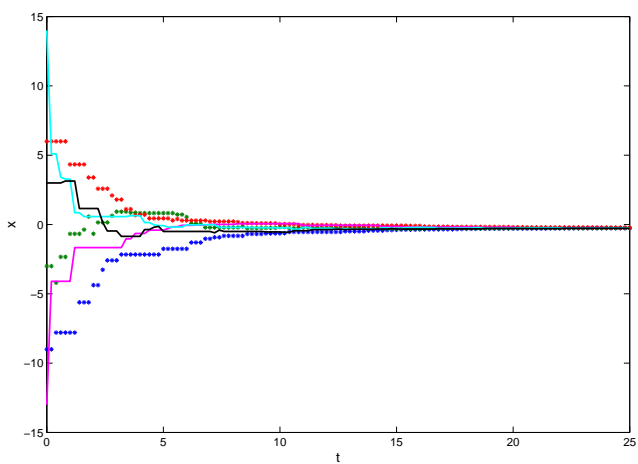}\\
  \caption{The state trajectories of all the agents with consensus protocol (\ref{gossip-CC})--(\ref{gossip-lf}) and network $\mathscr{G}_3$.}\label{result-ex3}
\end{figure}

\begin{example}\label{ex3}
Suppose that hybrid multi-agent system (\ref{mod-hybrid}) runs with control inputs (\ref{gossip-CC})--(\ref{gossip-lf}). The communication network is depicted in Fig. \ref{graph-ex3}. It is shown that the sampling period $h<\frac{1}{\max_{i,j=1}^6\{a_{ij}\}}$. At time $t_k$, each edge $(s_i,s_j)\in \mathscr{E}$ is chosen with probability $\frac{1}{7}$. The state trajectories of all the agents are drawn in Fig \ref{result-ex3}. Obviously, the simulation results is consistent with  the sufficiency of Theorem \ref{consensus-gossip}.
\end{example}

\section{Conclusions}\label{s-Conclusion}

In this paper, the consensus problem of hybrid multi-agent system which is composed of continuous-time and discrete-time dynamic agents was considered. First, we assumed that  all agents communicate with their neighbours and update their strategies in the sampling time. When the sampling period $0<h<\frac{1}{\max_{i\in \mathcal{I}_n}\{d_{ii}\}}$, we proved that the hybrid multi-agent system can achieve the consensus if and only if the communication network has a directed spanning tree. Then, we further assumed that each continuous-time agent can observe its own state in real time. The consensus of hybrid multi-agent system can be solved with $0<h<\frac{1}{\max_{i\in\mathcal{I}_{n}/\mathcal{I}_{m}}\{d_{ii}\}}$. Finally, a gossip-like consensus protocol was proposed. The necessary and sufficient condition was also given for solving the consensus problem if $0<h<\frac{1}{\max_{i,j\in \mathcal{I}_n}\{a_{ij}\}}$. The future work will focus on the second-order consensus of hybrid multi-agent system, consensus of hybrid multi-agent system with time-delays etc.

\QEDA



\ifCLASSOPTIONcaptionsoff
  \newpage
\fi



%

\end{document}